\documentstyle[12pt]{article}
\newcommand{\del}{{\bf \Delta}}
\newcommand{\dl}{{\delta{\cal L}}}
\newcommand{\delv}{{\bf \del}}
\newcommand{\msb}{{\overline{\rm MS}}}
\newcommand{\delfour}{{\Delta^{(4)}}}
\newcommand{\delsq}{\Delta^{(2)}}
\newcommand{\xv}{\vec{x}}
\newcommand{\yv}{\vec{y}}
\newcommand{\nscsk}{n_{sc},n_{sk}}
\newcommand{\tminmax}{t_{min}/t_{max}}

\newcommand{\zeroE}{{\cal E_0}}
\newcommand{\ainv}{{a^{-1}}}

\newcommand{\binding}{{\rm binding}}
\newcommand{\half}{{\mbox{$\frac{1}{2}$}}}
\newcommand{\be}{\begin{equation}}
\newcommand{\ee}{\end{equation}}
\newcommand{\order}{{\cal O}}
\newcommand{\lag}{{\cal L}}
\newcommand{\nr}{{\rm NR}}
\newcommand{\pv}{{\bf p}}
\newcommand{\mbare}{\mbox{$M_b^0$}}
\newcommand{\fig}[1]{Fig.~\ref{#1}}
\newcommand{\Ref}[1]{Ref.~\cite{#1}}
\newcommand{\kin}{{\rm kin}}
\newcommand{\eq}[1]{Eq.~(\ref{#1})}
\newcommand{\hfrac}[2]{{{#1}/{#2}}}
\newcommand{\nl}{\nonumber \\}
\newcommand{\Dv}{{\bf D}}
\newcommand{\Ev}{{\bf E}}
\newcommand{\Bv}{{\bf B}}
\newcommand{\psid}{{\psi^\dagger}}
\newcommand{\sigmav}{\mbox{\boldmath$\sigma$}}
\newcommand{\NRQCDcoll}{
C.~T.~H.~Davies,$^a$
K.~Hornbostel,$^b$
A.~Langnau,$^c$
G.~P.~Lepage,$^c$ \\
A.~Lidsey,$^a$
J.~Shigemitsu,$^d$
J.~Sloan$^e$ \\[.4cm]

\small $^a$University of Glasgow, Glasgow, UK G12 8QQ. \\
\small $^b$Southern Methodist University, Dallas, TX 75275. \\
\small $^c$Newman Laboratory of Nuclear Studies, Cornell University,
Ithaca, NY 14853. \\
\small $^d$The Ohio State University, Columbus, OH 43210. \\
\small $^e$Florida State University, SCRI, Tallahassee, FL 32306.
}

\begin{document}

\title{ Precision  $\Upsilon$ Spectroscopy from Nonrelativistic Lattice QCD}

\author{
\NRQCDcoll \\ }

\maketitle

\begin{abstract}
\noindent
The spectrum of the $\Upsilon$ system is investigated using the
Nonrelativistic Lattice QCD approach to heavy quarks and ignoring
light quark vacuum polarization. We find good agreement with experiment
 for the $\Upsilon$,
$\Upsilon^{\prime}$, $\Upsilon^{\prime\prime}$
and for the center of mass and fine
structure of the $\chi_b$-states.
The lattice calculations predict $b \bar{b}$ D-states with center of mass
at (10.20 $\pm$ 0.07 $\pm$ 0.03)GeV.
 Fitting procedures aimed at extracting both
ground and excited state energies are developed. We calculate a
nonperturbative dispersion
mass for the $\Upsilon(1S)$ and compare with
 tadpole-improved lattice perturbation theory.
\end{abstract}

\section{Introduction}
Hadrons containing one or more heavy quarks have been the focus of intense
 investigations by lattice gauge theorists in recent years. One motivating
 factor is that these systems are also being studied extensively by
experimentalists trying to nail down the remaining parameters in the Standard
Model.  Nonperturbative QCD results are needed in many instances, to convert
experimental numbers into determinations of fundamental parameters or to
test the Standard Model.  The lattice approach to nonperturbative QCD is
now starting to yield reliable numbers for several of these crucial inputs.
Part of the activity has been in heavy-light systems, focusing on leptonic
and semi-leptonic decays of heavy-light mesons (the B's and D's) and on
neutral meson mixing \cite{h-light}.
  Another area of investigation, which is also the focus
of the present article, has concentrated on heavy-heavy systems such as
the $J/\Psi$ and $\Upsilon$ families.  Studies of the latter systems have
already lead to the most accurate lattice determinations of the strong
coupling constant, $\alpha_s$ \cite{alpha,poster,wolff},
and of the b-quark pole mass $M_b$ \cite{M_b}.
In heavy-heavy
systems one can take advantage of the fact that only heavy quark propagators
are required to do high statistics simulations  at only modest computational
cost (of course only once the gauge configurations have been created).
 This coupled with the wealth of experimental data on quarkonium allows
one to carry out
 stringent tests of computational methods employed by lattice
gauge theorists.  These systems may also be the place where effects
of quenching can be studied quantitatively.
 We report here on a study of the $\Upsilon$ system using the Nonrelativistic
QCD (NRQCD) \cite{nrqcd,cornell}
 approach to heavy fermions. Our goal is to start from a first
principles QCD hamiltonian and show that it can reproduce the $\Upsilon$
spectrum.  Along the way we
 develop and refine methods to analyze numerical data, methods that we
hope will be useful in other lattice calculations as well.
For instance, we find that extracting excited state energies is
straightforward using our fitting procedures.
 Our simulations also serve
 to test perturbation theory on the lattice in a new
setting, through comparisons of nonperturbative simulation results for the
$\Upsilon$ kinetic mass with perturbative formulas.  The investigations
in this article provide the foundations for our determination of
the b-quark pole mass.   The $M_b$ calculations
are described in a separate publication\cite{M_b}.  We emphasize that
NRQCD provides an extremely efficient way to obtain realistic and accurate
heavy quark propagators.   The prospect of NRQCD having impact
not only in investigations of heavy-heavy, but also of heavy-light systems
looks very promising \cite{hlight}.

\vspace{.1in}

 The b-quarks in the $\Upsilon$ system are
quite nonrelativistic, with $v^2 \sim 0.1$.  The
 splittings between spin averaged levels are around $\sim$500 MeV
 ($\order(M_b \, v^2)$), which is much smaller than the mass ($\order(2M_b)$),
 indicating that a systematic expansion
of the QCD hamiltonian in powers of $v^2$ is very appropriate here.
The continuum action density, correct through $\order(M_b \, v^4)$,
 is given by

\be
\lag_{cont.} = \psid \left(D_t + H^{cont.}_0\right) \psi
 \; + \; \psid \, \delta H^{cont.} \, \psi
\ee
with

\begin{eqnarray}
H_0^{cont.} &=&  - \frac{\Dv^2}{2\mbare} \nl
\delta H^{cont.}&=& -\;\; c_1\,\frac{1}{8(\mbare)^3} \, (\Dv^2)^2
\;\; +\;\;  c_2\,\frac{ig}{8(\mbare)^2}\,  (\Dv\cdot\Ev - \Ev\cdot\Dv)  \nl
& & -\;\; c_3\,\frac{g}{8(\mbare)^2}\, \sigmav\cdot(\Dv\times\Ev -
\Ev\times\Dv)
\;\; -\;\;  c_4\,\frac{g}{2\mbare}\, \sigmav\cdot\Bv  .
\end{eqnarray}
$\psi$, $\psi^\dagger$ are two component Pauli spinors and
at tree-level we have $c_i = 1$ for all $i$. Previous NRQCD studies
\cite{nrqcd,bethchris} have
used the leading order Hamiltonian ($\delta H = 0$) or the leading order
plus the $\sigmav \cdot \Bv$ term and gave encouraging results.  Here we
include all the $\order(M_b \, v^4)$ terms.  This means systematic errors
due to relativity will be of $\order(M_bv^6)\, \sim \, 5 MeV$, which is
about 1\% of a typical radial or orbital excitation energy and 10\% of a
typical spin splitting.

In our lattice simulations there will be other sources of systematic errors.
Finite volume errors are not a problem here since the $\Upsilon$'s are
smaller than regular light hadrons.  Finite lattice spacing errors in
the fermionic action can be corrected for order by order, similar to
the systematic $v^2$ corrections.  So the main remaining sources of systematic
errors come from the gauge configurations, namely $\order(a^2)$ errors from
using the standard Wilson gauge action
and quenching errors due to the absence of light quark vacuum polarization.
{}From potential model calculations we estimate the latter errors to be
the dominant ones.  One consequence of quenching will be that inverse
lattice spacings, $a^{-1}$, extracted from different observables will
not agree with each other in general and one will have to
make some choice when presenting dimensionful results.
We find that our two basic splittings, the 1S-1P and the 1S-2S
$\Upsilon$ energy level splittings,  give $a^{-1}$'s that differ by one
to two $\sigma$.  We use an ``average'' lattice spacing in our
dimensionful plots.

\vspace{.1in}
Finally we need to discuss the number of  parameters in the NRQCD action.  In
addition to the bare mass, $\mbare$, and the gauge coupling, g, one has
the $c_i$'s.  We work with the $c_i$'s set to their tree-level values
 $c_i = 1$ while at the same time ``tadpole-improving'' the lattice version
of the NRQCD action
 \cite{impert}.  This ensures an optimal
perturbative scheme so that one can expect renormalization effects
to be small.  We find that in practice, tree-level values give the
correct P-state fine structure splittings, giving us confidence that
setting $c_i = 1$ is a viable approach.
The final parameter we must consider is the zero of energy which
is used to relate NRQCD energies with absolute, relativistic energies (this
term is usually omitted from  NRQCD actions).
As with the $c_i$'s, it can be fixed at tree-level to equal $\mbare$,
or calculated in perturbation theory.  It can also be
calculated non-perturbatively by requiring that the dispersion relation
of the $\Upsilon$ be Lorentz invariant, up to the order in $v^2$ at which
we are working.  We find excellent agreement between perturbative and
non-perturbative determinations, further encouraging us
that perturbation theory is working.  We stress that the only
free parameters which we tune to match experimental results are those
appearing in the original QCD action, $\mbare$ and g.
  In other words, this is a first-principles QCD
calculation, {\it not} a QCD inspired phenomenological model.  The coupling
g is eliminated as a free parameter, in the usual way, when we fix the
scale $a^{-1} = a^{-1}(g)$ to match the 1S - 1P and/or the  1S - 2S splitting.
$\mbare$ is tuned so that the simulated kinetic mass for the $\Upsilon$ agrees
with the experimental $\Upsilon(1S)$ mass.

\begin{figure}
\begin{center}
\setlength{\unitlength}{.02in}
\begin{picture}(80,140)(0,930)
\put(15,940){\line(0,1){120}}
\multiput(13,950)(0,50){3}{\line(1,0){4}}
\multiput(14,950)(0,10){10}{\line(1,0){2}}
\put(12,950){\makebox(0,0)[r]{9.5}}
\put(12,1000){\makebox(0,0)[r]{10.0}}
\put(12,1050){\makebox(0,0)[r]{10.5}}
\put(12,1060){\makebox(0,0)[r]{GeV}}
\put(25,940){\makebox(0,0)[t]{${^1S}_0$}}
\put(25,943){\circle*{2}}

\put(50,940){\makebox(0,0)[t]{${^3S}_1$}}
\multiput(43,946)(3,0){5}{\line(1,0){2}}
\put(50,946){\circle*{2}}

\multiput(43,1002)(3,0){5}{\line(1,0){2}}
\put(50,1004){\circle*{2}}
\put(50,1004){\line(0,1){3}}
\put(50,1004){\line(0,-1){3}}

\multiput(43,1036)(3,0){5}{\line(1,0){2}}
\put(50,1034){\circle*{2}}
\put(50,1034){\line(0,1){12}}
\put(50,1034){\line(0,-1){12}}

\put(75,940){\makebox(0,0)[t]{${^1P}_1$}}

\multiput(68,990)(3,0){5}{\line(1,0){2}}
\put(75,987){\circle*{2}}
\put(75,987){\line(0,1){2}}
\put(75,987){\line(0,-1){2}}

\multiput(68,1026)(3,0){5}{\line(1,0){2}}
\put(75,1032){\circle*{2}}
\put(75,1032){\line(0,1){7}}
\put(75,1032){\line(0,-1){7}}

\put(100,940){\makebox(0,0)[t]{${^1D}_2$}}
\put(100,1020){\circle*{2}}
\put(100,1020){\line(0,1){7}}
\put(100,1020){\line(0,-1){7}}

\end{picture}
\end{center}
 \caption{NRQCD simulation results for the spectrum of the
$\Upsilon$ system including radial excitations.
  Experimental
 values (dashed lines) are indicated for the triplet
$S$-states, and for the
 spin-average of the triplet $P$-states. The energy zero from
 simulation results is adjusted to give the correct mass to the
 $\Upsilon(1{^3S}_1)$.}
\label{spect}
\end{figure}
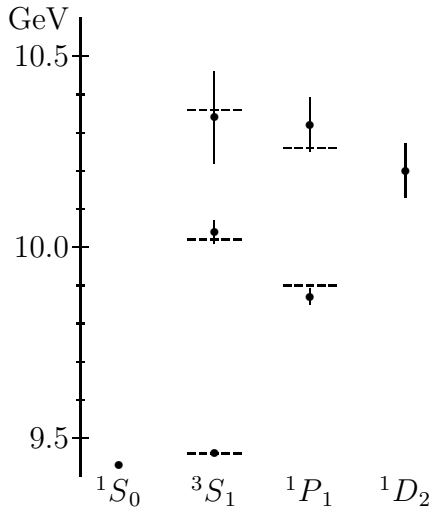

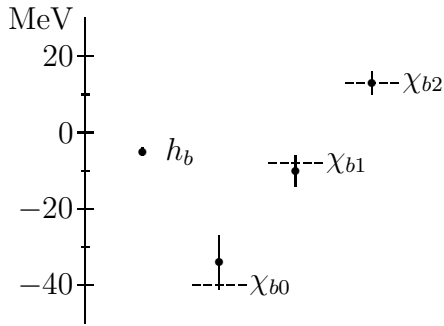
\begin{figure}
\begin{center}
\setlength{\unitlength}{0.02in}
\begin{picture}(110,140)(0,-50)
\put(15,-50){\line(0,1){80}}
\multiput(13,-40)(0,20){4}{\line(1,0){4}}
\multiput(14,-40)(0,10){7}{\line(1,0){2}}
\put(12,-40){\makebox(0,0)[r]{$-40$}}
\put(12,-20){\makebox(0,0)[r]{$-20$}}
\put(12,0){\makebox(0,0)[r]{$0$}}
\put(12,20){\makebox(0,0)[r]{$20$}}
\put(12,30){\makebox(0,0)[r]{MeV}}

\multiput(43,-40)(3,0){5}{\line(1,0){2}}
\put(58,-40){\makebox(0,0)[l]{$\chi_{b0}$}}
\put(50,-34){\circle*{2}}
\put(50,-34){\line(0,1){7}}
\put(50,-34){\line(0,-1){7}}

\put(36,-5){\makebox(0,0)[l]{$h_b$}}
\put(30,-5){\circle*{2}}
\put(30,-5){\line(0,1){1}}
\put(30,-5){\line(0,-1){1}}

\multiput(63,-8)(3,0){5}{\line(1,0){2}}
\put(78,-8){\makebox(0,0)[l]{$\chi_{b1}$}}
\put(70,-10){\circle*{2}}
\put(70,-10){\line(0,1){4}}
\put(70,-10){\line(0,-1){4}}

\multiput(83,13)(3,0){5}{\line(1,0){2}}
\put(98,13){\makebox(0,0)[l]{$\chi_{b2}$}}
\put(90,13){\circle*{2}}
\put(90,13){\line(0,1){3}}
\put(90,13){\line(0,-1){3}}
\end{picture}
\end{center}
 \caption{Simulation results for the spin structure of the lowest lying
 $P$-wave states in the $\Upsilon$~family. The dashed lines are
 the experimental values for the triplet states.  Energies
  are measured relative to the center of  mass of the triplet states.}
\label{fs}
\end{figure}

\vspace{.1in}
Our results for the spectrum of the $\Upsilon$ system are shown in
Fig.~\ref{spect} and Fig.~\ref{fs}.  We use
$a^{-1} = 2.4GeV$, which is an average between the inverse lattice spacings
obtained from the $\Upsilon$ 1S-2S and 1S-1P splittings (the error in this
estimate for $a^{-1}$
is at the 4\% level, details are given in section 4).
One sees that the general features of
the known spectrum are reproduced nicely \cite{previous}.
 Fig.~\ref{spect} shows the $\Upsilon$,
 $\eta_b$ and singlet P- and D- states and
Fig.~\ref{fs} shows the P-state fine structure.
In both figures the errors reflect statistical errors plus some systematic
fitting errors.  We do not show estimates of systematic
errors due to quenching or the effects of uncertainties in the scale $a^{-1}$.

\vspace{.1in}
  Both the $\eta_b$  and the D-state (center of mass)  are
predictions of the theory.
We find an $\eta_b$ state at (9.431 $\pm$
0.005 $\pm$ 0.001)GeV and D-states with center of mass at (10.20 $\pm$ 0.07
 $\pm$ 0.03)GeV.  These numbers include the dominant statistical
and/or systematic errors other than those due to quenching.  For the D-states
the first error corresponds to the statistical error in fitting the
D mass in lattice units.
  For the $\eta_b$ state this error is negligible and so the first error
quoted there is the systematic error from neglected higher order
relativistic terms and finite lattice spacing corrections.  In both cases
the second error arises from the uncertainty in the value of $a^{-1}$.
We expect the S-states
and hence also the $\Upsilon$ - $\eta_b$ splitting to have noticeable
quenching errors.  Spectrum calculations with dynamical gauge configurations
are already
underway.  It will be interesting  to compare the quenched
and unquenched spectra.

\vspace{.1in}
In the rest of the article we give more details of our analyses, starting
with the quark propagator calculations
and the meson correlations
in the next section. Section 3
includes a lengthy explanation of our fitting procedures.  The main message
there is that, instead of going to large timeslices in search of a plateau,
we have worked with high statistics correlations on shorter lattices
(24 time slices) and carried out simultaneous multi-exponential fits
to several correlations at a time.  We also devised methods to extract
splittings directly.  Our statistical errors for spin averaged level
splittings are about 20 - 30 MeV, for P-state fine structure splittings
about 5 - 10 MeV and for the S-state spin splitting down to 0.5MeV.
In section 4 we discuss determination of $a^{-1}$ and comparison with
experiment.  We also present results for meson wave functions at the
origin.
Section 5 describes the extraction of a ``kinetic mass'', $aM_{kin}$,
for the $\Upsilon(1S)$ from correlations with momenta and comparisons
 with perturbation theory.
Section 6 gives a brief summary.

\vspace{.2in}
\section {The Simulation}

\subsection{Quark Propagators}
Quark propagators in lattice NRQCD are determined, in a single pass
through the gauge-field configuration, from evolution equations that
specify the propagator for $t>0$ in terms of its value at $t=0$.
Various evolution equations have been suggested in the past. Currently
we use the equation proposed
in~\cite{cornell}, modified slightly for improved efficiency.
Our propagators are defined by the equation
\be \label{tevolve-nl}
\left(1\!-\!\frac{aH_0}{2n}\right)^{-n} \,
U_4 \,
\left(1\!-\!\frac{aH_0}{2n}\right)^{-n} \,
G_{t+1}\, - \,
\left(1\!-\!a\,\delta H\right)
 G_t  = \delta_{\xv,0}\,\delta_{t,0}
 \ee
where $G_t = 0$ for $t\le 0$. For numerical work it is convenient to
rewrite this equation in the form:
\begin{eqnarray}
 G_1 &=&
  \left(1\!-\!\frac{aH_0}{2n}\right)^{n}
 U^\dagger_4
 \left(1\!-\!\frac{aH_0}{2n}\right)^{n} \, \delta_{\xv,0}  \nl
  G_{t+1} &=&
  \left(1\!-\!\frac{aH_0}{2n}\right)^{n}
 U^\dagger_4
 \left(1\!-\!\frac{aH_0}{2n}\right)^{n}\left(1\!-\!a\delta H\right) G_t
 \quad (t>1) .
\label{tevolve}
\end{eqnarray}
On the lattice, the kinetic energy operator is
 \be
 H_0 = - {\delsq\over2\mbare},
 \ee
and the correction terms are
 \begin{eqnarray}
\delta H
&=& - c_1 \frac{(\delsq)^2}{8(\mbare)^3}
            + c_2 \frac{ig}{8(\mbare)^2}\left(\delv\cdot\Ev -
\Ev\cdot\delv\right) \nl
 & & - c_3 \frac{g}{8(\mbare)^2} \sigmav\cdot(\delv\times\Ev -
\Ev\times\delv)
 - c_4 \frac{g}{2\mbare}\,\sigmav\cdot\Bv  \nl
 & &  + c_5 \frac{a^2\delfour}{24\mbare}
     -  c_6 \frac{a(\delsq)^2}{16n(\mbare)^2} .
\end{eqnarray}
The last two terms in $\delta H$ come from finite lattice spacing
corrections to the lattice laplacian and the lattice time derivative
respectively.  $\delv$ is the symmetric lattice derivative and
 $\delfour$ is a lattice version of the continuum
operator $\sum D_i^4$. We used the standard cloverleaf operators for the
chromo-electric and magnetic fields, $\Ev$ and~$\Bv$.  The
parameter~$n$
is introduced to remove instabilities in the heavy quark
propagator caused by the highest momentum modes of the theory.
For our simulations at $\beta = 6.0$ and with bare masses relevant for the
$\Upsilon$ system, we set $n = 2$.

 As mentioned in the Introduction, we tadpole-improve our lattice
action by dividing all the $U's$ that appear in $\Ev$, $\Bv$, and
 the covariant lattice derivatives
fields by $u_0$, the fourth root of the plaquette\cite{impert}.
This is most easily done as the $U_\mu$'s
are read by the simulation code.  The effect is to transform the
operators that appear in the evolution equations~(\ref{tevolve})
as follows:
\begin{eqnarray}
U_4^\dagger & \longrightarrow & U_4^\dagger/u_0  \\
\Ev &\longrightarrow & \Ev/(u_0)^4 \\
\Bv &\longrightarrow & \Bv/(u_0)^4 \\
\Delta_\mu G(x) &  \longrightarrow & \left( U_\mu(x) G(x+\hat\mu) -
   U^\dagger_\mu(x-\hat\mu) G(x-\hat\mu) \right)/2u_0 \\
\Delta_\mu^{(2)}\, G(x) &
   \longrightarrow & \left( U_\mu(x) G(x+\hat{\mu})
            + U^\dagger_\mu(x-\hat{\mu}) G(x-\hat{\mu}) \right)/u_0
                      - 2 G(x)
\end{eqnarray}
with
\begin{eqnarray}
\delsq  &=& \sum_\mu  \Delta_\mu^{(2)}  \\
\delfour&=& \sum_\mu (\Delta_\mu^{(2)})^2
\end{eqnarray}
Tadpole-improvement of the action allows us to
work with tree-level values for the $c_i$'s in $\delta H$ without
having to worry about large renormalizations. Hence our lattice action
depends only on two parameters, the bare mass~$\mbare$ and the QCD coupling
constant~$g$.
We have collected data for three values of the bare mass, $a\mbare$ =
1.71, 1.8 and 2.0, all at $\beta$ = 6.0.  For this~$\beta$, $u_0 = 0.878$.

We computed our quark propagators for $a\mbare=1.8,\,2.0$ using a different
evolution equation:
\be
U_4\,G_{t+1} -
\left(1\!-\!\frac{aH_0}{2n}\right)^{n}
\left(1\!-\!a\delta H\right)
 \left(1\!-\!\frac{aH_0}{2n}\right)^{n} \, G_t = \delta_{\xv,0}
\,\delta_{t,0}.
\label{nevolve}
\ee
Propagators from this evolution equation are the same as those from
the other (\eq{tevolve-nl}), but with extra factors of
$\left(1-aH_0/2n\right)^{-n}$ in both
the source and sink. These extra factors cannot affect the meson
spectrum, but they do modify the quark's wavefunction normalization in
order~$v^2$. For this reason our other formulation (\eq{tevolve-nl})
is superior; with it, NRQCD quark fields are normalized to unity up to
corrections of order $\alpha_s$ and $v^4$.

\vspace{.3in}
\subsection{ Meson Correlation Functions }
  Once one has the quark propagators it is straightforward to obtain
meson propagators.  Let $\psi^\dagger$ and $\chi^\dagger$ denote
fields that create a heavy quark or heavy anti-quark respectively.
The following interpolating operator creates a
meson of momentum $\vec{p}$.

\be
 \sum_{\xv_1,\xv_2} \psid(\xv_1) \, \Gamma(\xv_1 - \xv_2) \, \chi^\dagger
(\xv_2)\,e^{i{\vec{p} \over 2}\cdot (\xv_1 + \xv_2) }
\label{meson}
\ee
where the ``meson operator'' =
 $\Gamma(\xv_1 - \xv_2) = \Omega \; \phi(|\xv_1 - \xv_2|)$.  The operator
$\Omega$ is a $2 \times 2$ matrix in spin space and generally includes
derivatives acting on the radial function $\phi(r)$.  Using translation
invariance, we eliminate the summation over the initial antiquark position.
The meson propagator is then

\be
 G_{meson}(\vec{p},t) =
 \sum_{\yv_1,\yv_2}Tr \left[ G^\dagger_t(\yv_2) \,
{\Gamma^{(sk)}}^\dagger (\yv_1 - \yv_2) \,
\tilde{G}_t(\yv_1) \right] e^{-i{\vec{p} \over 2} \cdot (\yv_1 + \yv_2)}
\label{gmes}
\ee
with
\be
\tilde{G}_t(\yv) \equiv \sum_{\xv}G_t(\yv - \xv) \, \Gamma^{(sc)}(\xv)
e^{i {\vec{p} \over 2} \cdot\xv}
\ee
and the trace is over spin and color.  In the above equations we
distinguish between $\Gamma^{(sc)}$ and $\Gamma^{(sk)}$, i.e. the smearing
at the source or sink.
 $\tilde{G}_t(\xv)$
can be obtained directly using \eq{tevolve} with $ \delta_{\xv,
\vec{0}} \rightarrow  \Gamma^{(sc)}(\xv) \, e^{i
 {\vec{p} \over 2}
\cdot \xv}$.  In the future we will often refer to the smeared propagator
$\tilde G$ as the quark propagator.
The convolution in \eq{gmes} is evaluated
using fast fourier transforms.

\begin{table}
\begin{center}
\begin{tabular}{cccc}
Meson &Lattice  & $\Gamma = \Omega \phi (r)$ & \\
$^{2S+1}L_J$ ($J^{PC}$)&  Rep.  & $\Omega$& $\phi$ \\
\cline{1-4}
&&& \\
 ${^1S}_0\;(0^{-+})$ &  $A^{-+}_1$  &$\hat{I}$ &$\phi^{(S)}_{n_{sc}}(r)$ ;
$\;\;n_{sc}$ = loc,1,2,3  \\
&&& \\
 ${^3S}_1\;(1^{--})$ &  $T^{--}_{1(i)}$  & $\sigma_i $ &   \\
&&& \\
 ${^1P}_1\;(1^{+-})$ &  $T^{+-}_{1(i)}$  & $\Delta_i $
                                    &$\phi^{(P)}_{n_{sc}}(r)$;
$\;\; n_{sc}$ = loc,1,2  \\
&&& \\
 ${^3P}_0\;(0^{++})$ &  $A^{++}_1$  & $\sum_j  \Delta_j  \sigma_j$ &  \\
&&& \\
 ${^3P}_1\;(1^{++})$ &  $T^{++}_{1(k)}$  & $ \Delta_i \sigma_j - \Delta_j
\sigma_i $ &  \\
&&& \\
 ${^3P}_2\;(2^{++})$ &  $E^{++}_{(k)}$  & $ \Delta_i \sigma_i - \Delta_j
\sigma_j$ & \\
&&& \\
                     &  $T^{++}_{2(ij)}$  & $ \Delta_i \sigma_j + \Delta_j
\sigma_i$ & \\
&&$\qquad$ ($i \neq j$) & \\
&&& \\
 ${^1D}_2\;(2^{-+})$ &  $E^{-+}_{(k)}$  & $ D_{ii} - D_{jj}$ &
$\phi^{(D)}_{n_{sc}}(r)$ ;
$\;\; n_{sc}$ = loc,1,2  \\
&&& \\
                   &  $T^{-+}_{2(ij)}$  & $ D_{ij}$  & \\
&&$\qquad$ ($i \neq j$) & \\
&&& \\
 ${^3D}_1\;(1^{--})$ &  $T^{--}_{1(i)}$  & $\sum_j D_{ij}  \sigma_j $ & \\
&&& \\
 ${^3D}_2\;(2^{--})$ &  $E^{--}_{(j)}$  & $ D_{ij} \sigma_k - D_{jk}
\sigma_i$ & \\
&&& \\
                &  $T^{--}_{2(ij)}$
                     & $ (D_{ii} - D_{jj}) \sigma_k \;$ + & \\
&&$D_{jk} \sigma_j - D_{ki} \sigma_i$ & \\
&&& \\
 ${^3D}_3\;(3^{--})$ &  $A^{--}_2$  & $ (D_{ij} \sigma_k + D_{jk}
\sigma_i + D_{ki} \sigma_j)/3$ & \\
&&& \\
                     &  $T^{--}_{1(i)}$  & $ D_{ii} \sigma_i
- 2/5 \sum_j D_{ij} \sigma_j $ & \\
&&& \\
                     &  $T^{--}_{2(ij)}$
            & $ (D_{ii} - D_{jj}) \sigma_k \;+ $ & \\
&&$2(D_{ki} \sigma_i - D_{kj} \sigma_j)$ & \\
\cline{1-4}

\end{tabular}
\end{center}
\caption{Meson Operators. $\Delta_i$ denotes the symmetric lattice
derivative and $D_{ij} \equiv \Delta_i \Delta_j - \delta_{ij}
\Delta^2/3$.}
\label{mesops}
\end{table}

\vspace{.2in}
In Table~\ref{mesops}  we list the zero momentum
$b \, \bar{b}$ meson states studied in the current
project together with their corresponding ``meson operators'',
$\Gamma(\xv)$.
 We choose $\phi_{n_{sc}}(r)$ for $n_{sc} = 1,2,3$ to correspond to
Richardson potential radial wave functions for the S-, P- or D-state
ground and excited levels.  We also used $\delta$-function
local sources and sinks.  These are referred to as $n_{sc}=loc$ in
Table~\ref{mesops}.
  For each set of meson quantum numbers we evaluated correlations
with all possible independent smearings,$(\nscsk) = (loc,loc), (loc,1)
, (loc,2) ...$ etc.,
 at the source and the sink.  So for the S-states we obtained a $4 \times 4$
matrix of correlation functions
and for the P- and D-states  $3 \times 3$ matrices.
In addition we looked at S-state mesons with small momenta.  For those we
used $\phi_{loc}$ and $\phi_1$.

\vspace{.2in}
 Table~\ref{mesops} also lists the continuum quantum numbers $J^{PC}$ and
the lattice cubic group representations for our meson states.  This allows
us to see which states are expected to mix with each other.  Mixing will
occur both due to relativistic corrections and due to lattice artifacts.
For instance, since $L$ is no longer a good quantum number in a relativistic
theory the ${^3S}_1$ and the ${^3D}_1$ states will mix, both being
$J^{PC} = 1^{--}$.  This mixing, which happens even in the
continuum limit, is suppressed by $v^2$.  On the lattice one can
also have mixing between different $J$-states that fall into the same lattice
representations.  Examples are mixing between ${^3D}_3T_2$ and
${^3D}_2T_2$ or between ${^3D}_3T_1$, ${^3D}_1T_1$ and ${^3S}_1$.
We have measured cross-correlations between these states, but postpone
their analysis for future work, concentrating here on the spin-averaged
D-states.

\vspace{.1in}
In the expression \eq{meson} one has sums over color and spin degrees of
freedom.  One could calculate quark (and antiquark) propagators
separately for each color and spin quantum number at the source.  We have
done so for the color degrees of freedom and verify a reduction of
statistical errors by $\sqrt{3}$ compared to when only one value for
the initial quark and antiquark color was used.
 As far as spin is concerned we decided
to save on CPU time by setting the initial quark and antiquark spins
equal to 1. This means that at the source we are sensitive only to the 1-1
component of the $2 \times 2 $ spin matrix in $\Gamma^{(sc)}$ and mesons of
definite quantum numbers are projected out at the sink.
{}From Table~\ref{mesops} one sees that
groups of mesons such as
$ {^1S}_0$ and $ {^3S}_1z$ , or
${^1P}_1x \; ,\; {^3P}_1y $ and $ {^3P}_2T_2zx$ etc.
 have the same 1-1 component of $\Gamma^{(sc)}$
up to normalization.
For each group,
the meson correlations for its members can
be obtained from one common quark propagator (this must be repeated
for each smearing function, $\phi_{n_{sc}}$ at the source), and are highly
correlated.  We have taken advantage of these strong correlations to
reduce statistical errors in our fits for hyperfine and fine structure
splittings.
We worked with 13 zero momentum quark propagators with
S- or P-state smearing at the source and four S-state propagators
 with momentum.  Out of these  quark propagators
129 S- and P-state meson correlations were evaluated
using different $\Gamma^{(sk)}$ and
many combinations of smearing functions at source and sink.  For $a\mbare =
1.71$ we also evaluated an additional 15 quark propagators with D-state
smearing. To date we have only analyzed combinations of these D-state
quark propagators giving rise to the $^1{D_2}$ mesons.  With 5 polarizations
and 9 different source-sink smearing combinations, this means a total of
45 D-state meson correlations.

\vspace{.1in}
\section{Data Analysis and Fitting Results}
Our calculations used $\beta = 6.0$ quenched gauge field configurations on
$16^3 \times 24$ lattices provided by Greg Kilcup and his collaborators.
  We worked with an ensemble of
105 independent configurations that were gauge-fixed to Coulomb gauge.
 For each configuration we selected 8
different origins on the first time slice for our quark (antiquark)
propagators. Simple tests, in which data from different origins but
from the same configuration were binned,
 indicated that different origins lead to
independent propagators. For our $a\mbare = 1.71$ P-states we also
ran with 8 origins on timeslice 12.  This data was binned together
with the timeslice 1 data for each configuration and spatial origin.
  Hence for each of the 129 (or 129 + 45) meson correlations,
discussed above, we worked with $105 \times 8 = 840$ measurements.
  In Fig.3 and Fig.4 we
show some examples of effective mass plots for our data. The errors are
bootstrap errors.  Fig.3 shows $a\mbare = 1.8$
data for ${^1S}_0$ states.  The effective mass plots are arranged as
a $3 \times 3$ matrix corresponding to the 9 source-sink combinations
$(\nscsk)$ with $\nscsk = loc,1,2$.  Fig.4 shows similar plots for
$a\mbare = 1.71$  ${^1P}_1$ data.  We have averaged over ${^1P}_1x$,
${^1P}_1y$ and ${^1P}_1z$.  This $h_b$ state has not been observed yet
experimentally but there are strong theoretical reasons for believing
that it lies close to the center of mass
of the ${^3P}$ levels. Hence we will sometimes refer to the ${^1P}_1$
level as the ``spin averaged'' P-state.  One sees from Fig.3 and
Fig.4  that
our S-state correlations with $n_{sk} = loc$ and $n_{sk} = 1$ sinks
have excellent statistics.  The S-state correlations with excited
state smearing $n_{sk} = 2,3$ and the P-state data have good
to reasonable statistics.  In the effective mass plots, truncations
at large $t$ means that signal to noise in the original data
was worse than $3:1$ beyond that point
or that the correlation had switched sign.  We have used the naive
definition  $m_{eff}(t) = log(G(t)/G(t+1))$, although a more
sophisticated version could be used when one deals with off-diagonal
correlations in which some amplitudes can come in with negative signs.
Our plots provide a rough assessment of the quality of the data; we do not
use them in our fits.

\renewcommand{\Diamond}{}

\setlength{\unitlength}{0.240900pt}

\begin{figure}
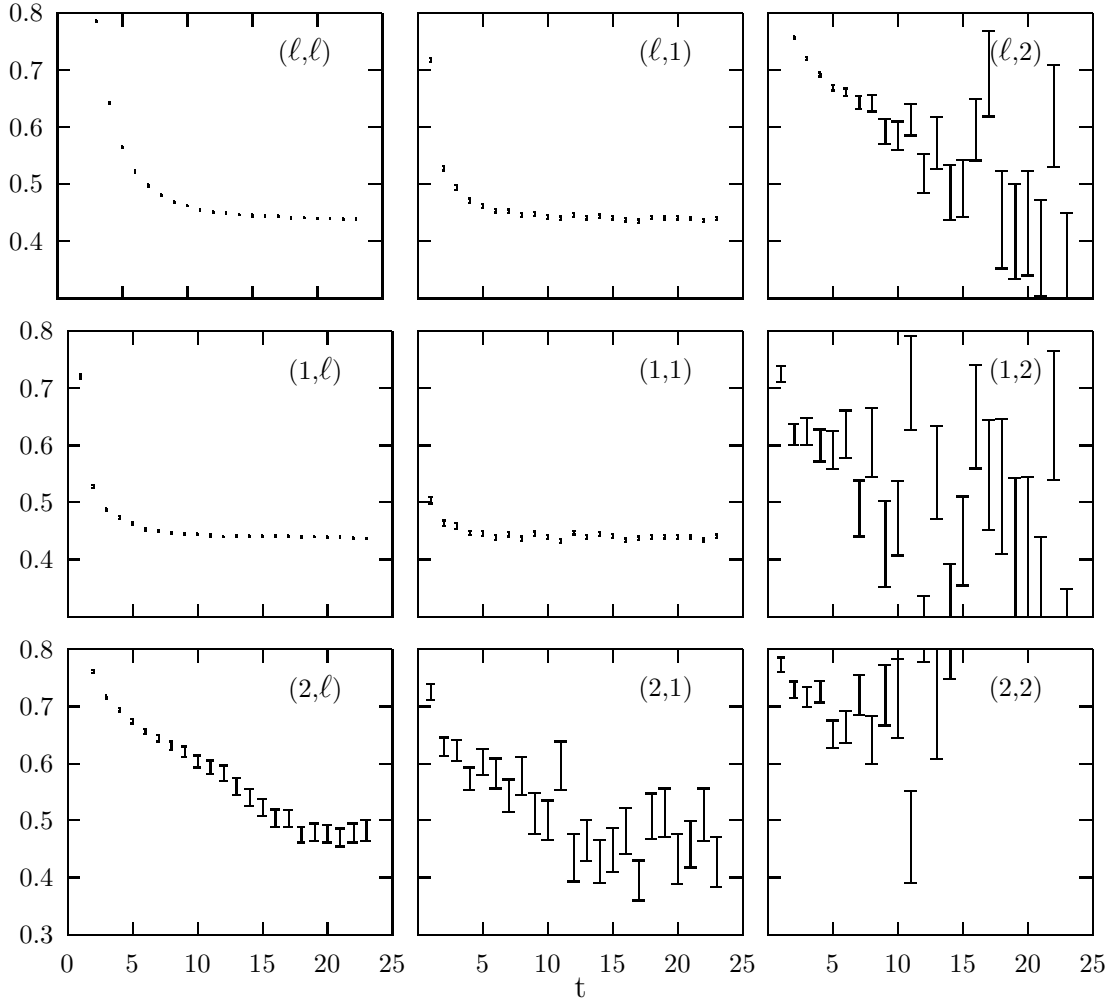


\noindent


\caption{$^1S_0$ Effective masses by (source, sink).}
\end{figure}

\renewcommand{\Diamond}{}

\setlength{\unitlength}{0.240900pt}

\begin{figure}
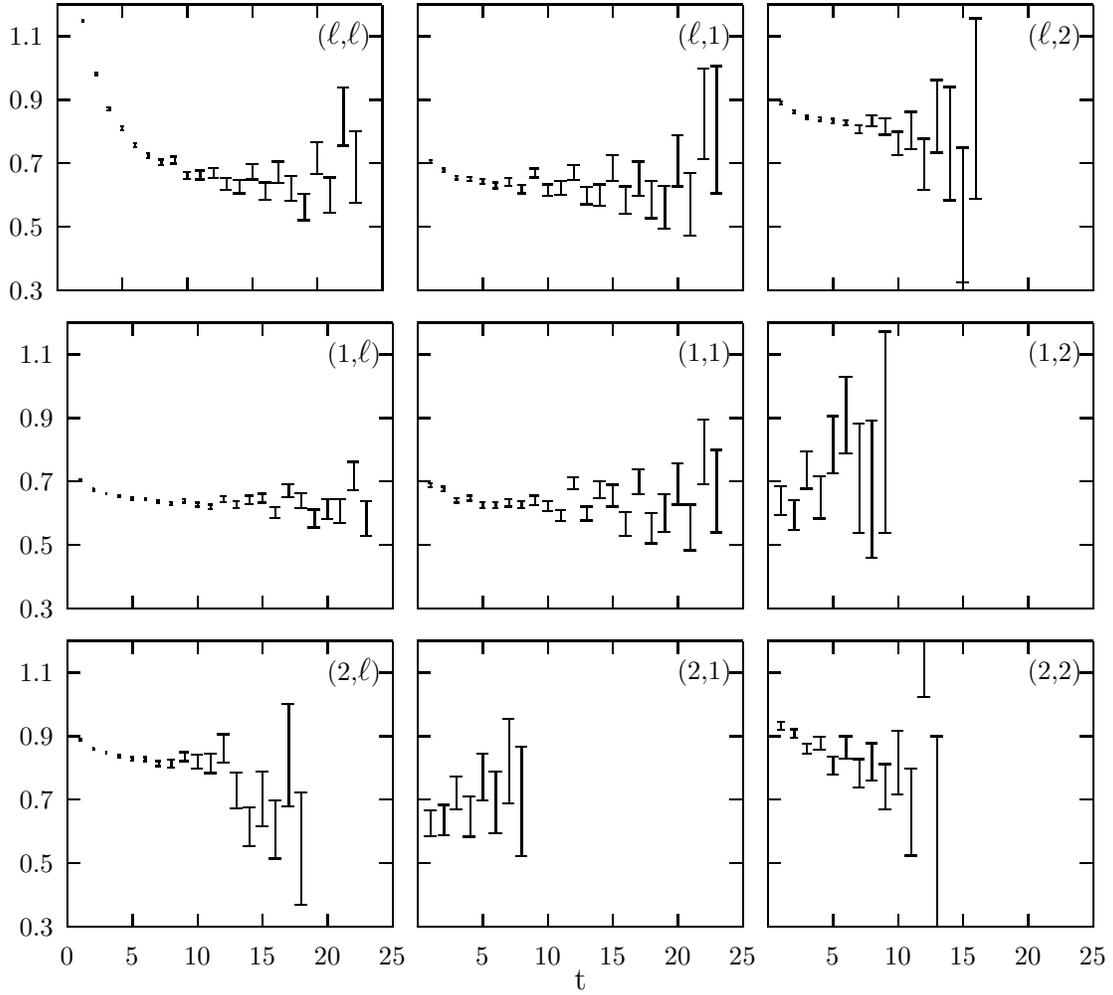


\noindent
%
}
\put (1005, -30) {t}
}
\end{picture}

\caption{$^1P_1$ Effective masses by (source, sink).}
\end{figure}

\vspace{.2in}
\subsection{Fitting Procedures for ${^3S}_1$ and Singlet P- and D-States}
We investigated a
variety of fitting procedures to extract the spectrum from the meson
correlations.  For the S-states and singlet P- and D-states we used
simultaneous multi-exponential fits to several correlations in order to
obtain the ground state and one or two excited state energies.  We
employed mainly two such fitting procedures and got consistent results
from both of them within one $\sigma$.  In the first procedure, we
fit simultaneously to a matrix of correlations with $\nscsk = 1,2,3$
for S-states and $\nscsk = 1,2$ for the spin singlet P-
 and D-states.  Similar fitting methods have been discussed by
other groups in the past\cite{matrixfit}.
In the matrix fits each meson correlation is fit to the
expression

\be
G_{meson}(\nscsk;t) = \sum_{k=1}^{N_{exp}}\; a(n_{sc},k)\,a^*(n_{sk},k)\,
e^{-E_k \cdot t} .
\label{matansatz}
\ee
Using charge conjugation symmetry one can argue that, for equal mass quarks,
the coefficients $a(j,k)$ can be chosen to be real numbers.
{}From the effective mass plots of Fig.3 and  Fig.4
it is clear that if one wants to
make use of most of the data
one does not want to move too far out with $t_{min}$ in search of
a plateau.  One wants instead to fit to several exponentials working with
as small a $t_{min}$ value as possible while still maintaining good
chi-squareds.
Multi-exponential fits to a single correlation, however
 are usually tricky and unstable.  It is much easier to do multi-exponential
fits to several correlations simultanously, especially if one makes sure that
each exponential has a significant amplitude in at least one correlation.
Although the number of parameters in our fits was sometimes large
(we were typically doing 6-,9-,12- and 16-parameter fits)
they were highly constrained and we did not run into stability problems.
Since we had a large number, 840, of measurements to play with we also did
not have to worry about the size of the covariance matrix becoming too
large.  We used $N_{exp} = 3$ for the $3 \times 3$ matrix fits and
$N_{exp} = 2$ for the $2 \times 2$ matrix fits.

\vspace{.1in}
The second procedure for doing multi-correlation multi-exponential fits
was to take a set of smeared-local
 correlations, $(\nscsk) = (n_{sc},loc)$ with
$n_{sc} = 1,2,3$ or $n_{sc} = 1,2$ and fit them simultaneously to
the same set of energies, $E_k$.  These smeared-local
 correlations have the smallest statistical
errors. They are fit to

\be
G_{meson}(n_{sc},loc;t) = \sum_{k=1}^{N_{exp}}\; b(n_{sc},k)\,
e^{-E_k \cdot t} .
\label{mcorfit}
\ee
Note that this is the same ansatz as \eq{matansatz}, with
$b(n_{sc},k) = a(n_{sc},k)a^*(loc,k)$.
We used $N_{exp} = 2 $ or $3$ when fitting two correlations and
$N_{exp} = 3 $ or $4$ for three correlations.  We use different
sets of correlations for fitting procedures 1 and 2.  So the fact that,
as we will see,
 consistent energies are obtained from the two procedures
gives us confidence in the final results.

\vspace{.1in}
In fitting correlations the delicate question is always how to pick
the range $\tminmax$ in $t$ over which to fit.  Needless to say, in our
multi-exponential fits the larger $N_{exp}$ the smaller we can make
$t_{min}$.  On the other hand for fixed $N_{exp}$ when one increases $t_{min}$
there could come a point when the signal for the higher exponential is no
longer above the noise leading to larger errors in the energies of even
the lower lying levels.   To
illustrate some of these issues we show in Table~\ref{smeared-local}
 examples of fits
to $a\mbare = 1.71$ ${^3S}_1$-states using procedure 2.
 We tabulate results for
different number of exponentials and correlations and show the dependence
 on $\tminmax$.  The last column gives Q, the ``goodness of
fit'', i.e. the probability that fluctuations in correctly modelled
data will generate a $\chi^2$ greater than that of the fit.
   One usually desires $Q > 0.1$, but slightly
smaller values do not necessarily rule out a fit.
Table~\ref{smeared-local} starts with
two exponential fits to two correlations with
$(\nscsk)$ = (1,loc) and (2,loc).
 One sees that for $t_{min}$ of 6 to 8 good fits are obtained to
the ground- and first excited level. For $t_{min} = 4$ the Q-value
starts to deteriorate.  Adding a third exponential  allows one
to go down to $t_{min}$ of 2. The third energy, $E_3$, however is not
reliable yet until we include another correlation with $\phi_3$ smearing.
Such fits, i.e. simultaneous fits to three correlations,
$G_{{^3S}_1}(1,loc;t)$, $G_{{^3S}_1}(2,loc;t)$ and
$G_{{^3S}_1}(3,loc;t)$ are also shown in Table~\ref{smeared-local}.

\begin{table}
\begin{center}
\begin{tabular}{l|cclllc}
& $N_{exp}$ & $t_{min}/t_{max}$&$\;\;aE_1 \;\;$&$\;\;aE_2\;\;$ &
 $\;\;aE_3\;\;$ & $Q$\\
\cline{1-7}
fits to (1,loc)& 2  & 8/24 & 0.4533(8) &0.707(9) & &0.62  \\
and (2,loc)    &    & 7/24 & 0.4531(8) &0.702(8) & &0.63  \\
               &    & 6/24 & 0.4531(7) &0.707(6) & &0.68 \\
               &    & 5/24 & 0.4533(7) &0.712(5) & &0.62  \\
               &    & 4/24 & 0.4533(7) &0.725(5) & & 0.04  \\
               &    & 3/24 & 0.4537(7) &0.740(4) & & $7 \times 10^{-5}$\\
               & 3  & 5/24 & 0.4531(8) &0.70(2)  &1.0(6) &0.57  \\
               &    & 4/24 & 0.4534(8) &0.69(1)  &1.1(2) &0.54  \\
               &    & 3/24 & 0.4531(4) &0.69(1)  &1.1(1) &0.48  \\
               &    & 2/24 & 0.4531(7) &0.69(1)  &1.0(1) &0.45  \\
\cline{1-7}
fits to (1,loc), & 3  & 7/24 & 0.4533(9) &0.72(3) & 0.82(5) &0.76 \\
(2,loc), (3,loc)&   & 6/24 & 0.4533(9) &0.72(5) & 0.83(8) &0.77 \\
                &  &  5/24 & 0.4530(8)& 0.71(7) & 0.85(9)& 0.75 \\
                &  &  4/24 & 0.4535(8)& 0.68(1) & 0.91(2)& 0.64 \\
                &  &  3/24 & 0.4531(8)& 0.69(1) & 0.95(1)& 0.33 \\
                &4 &  4/24 & 0.4529(9)& 0.71(4) & 0.82(10)& 0.74 \\
                &  &  3/24 & 0.4530(8)& 0.71(2) & 0.80(2)& 0.51 \\
\cline{1-7}
\end{tabular}
\end{center}
\caption{Examples of simultaneous multi-exponential fits to two
and three ${^3S}_1$ smeared-local correlations.}
\label{smeared-local}
\end{table}

\begin{table}
\begin{center}
\begin{tabular}{l|cclllc}
& $N_{exp}$ & $t_{min}/t_{max}$ &$\;\;aE_1\;\;$ &
 $\;\;aE_2\;\;$ & $\;\;aE_3\;\;$ & $Q$\\
\cline{1-7}
fits to      & 2  & 5/24 & 0.4537(7) &0.694(7) & &0.10  \\
(1,1), (1,2) &    & 5/22 & 0.4539(7) &0.693(7) & &0.17  \\
(2,1), (2,2) &    & 6/24 & 0.4536(7) &0.708(9) & &0.24  \\
             &    & 6/22 & 0.4538(7) &0.708(9) & &0.41  \\
             &    & 6/16 & 0.455(1) &0.71(1) & &0.21  \\
             &    & 7/24 & 0.4536(7) &0.70(1) & &0.17  \\
             &    & 7/22 & 0.4539(7) &0.70(1) & &0.31  \\
             &    & 8/24 & 0.4534(7) &0.71(1) & &0.13  \\
             &    & 8/22 & 0.4537(8) &0.71(1) & &0.24  \\
\cline{1-7}
fits to      & 3  & 4/24 & 0.4540(6) &0.702(6) &0.88(2) &0.04  \\
(1,1), (1,2) &    & 5/24 & 0.4538(6) &0.697(8) &0.86(3) &0.06  \\
(2,1), (2,2) &    & 5/22 & 0.4540(7) &0.696(8) &0.86(3) &0.06  \\
(1,3), (3,1) &    & 6/24 & 0.4536(7) &0.710(9) &0.90(7) &0.14  \\
(2,3), (3,2) &    & 6/22 & 0.4538(7) &0.709(9) &0.89(7) &0.18  \\
(3,3)        &    & 6/16 & 0.455(1) &0.711(9) &0.90(9) &0.05  \\
             &    & 7/24 & 0.4537(7) &0.67(3) &0.74(2) &0.22  \\
             &    & 7/22 & 0.4540(7) &0.67(3) &0.74(2) &0.31  \\
             &    & 8/24 & 0.4534(7) &0.64(5) &0.724(9) &0.49  \\
\cline{1-7}
\end{tabular}
\end{center}
\caption{Examples of $2 \times 2$ and $3 \times 3$ matrix fits to
 ${^3S}_1$ correlations.}
\label{matrix}
\end{table}

\begin{table}
\begin{center}
\begin{tabular}{l|cclll}
 $\qquad$ Fit $\qquad$& $t_{min}/t_{max}$ & k &$a(n_{sc,sk}=1,k)$ &
 $a(n_{sc,sk}=2,k)$ & $a(n_{sc,sk}=3,k)$\\
\cline{1-6}
$N_{exp} = 2$ fits to&6/22 &1& $\;\,$0.877(5) & -0.065(3)  &      \\
(1,1), (2,2)         &     &2& $\;\,$0.18(1)  &$\;\,$0.79(3)&     \\
\cline{2-6}
(1,2), (2,1)         &7/22 &1& $\;\,$0.878(5) &-0.066(3)   &      \\
                     &     &2& $\;\,$0.18(2)  &$\;\,$0.78(3)&     \\
\cline{1-6}
$N_{exp} = 3$ fits to&6/22 &1& $\;\,$0.877(5) &-0.065(3)&-0.015(1)\\
(1,1), (1,2), (2,1)  &     &2& $\;\,$0.18(1)  &$\;\,$0.79(2)&
$\;\,$0.07(4)  \\
(1,3), (3,1), (2,2)  &     &3& $\;\,$0.094(9) &-0.1(1) & $\;\,$0.8(2)\\
\cline{2-6}
(3,3)                &5/22 &1&$\;\,$0.878(4)  &-0.066(3)&-0.014(1) \\
                     &     &2& $\;\,$0.17(1)  &$\;\,$0.75(2)&
$\;\,$0.03(4)  \\
                     &     &3& $\;\,$0.09(1) &$\;\,$0.09(8)&$\;\,$0.70(6)\\
\cline{1-6}
\end{tabular}
\end{center}
\caption{Examples of fit results for amplitudes $a(n_{sc,sk},k)$}
\label{aamplitude}
\end{table}

\begin{table}
\begin{center}
\begin{tabular}{l|cclll}
 $\qquad$ Fit $\qquad$& $t_{min}/t_{max}$ & k &$b(n_{sc}=1,k)$ &
 $b(n_{sc}=2,k)$ & $b(n_{sc}=3,k)$\\
\cline{1-6}
$N_{exp} = 2$ fits to& 6/24 & 1 &0.135(2) &-0.0102(8) &  \\
 (1,loc) and (2,loc)&        & 2 &0.032(4) &$\;\,$0.114(2)&  \\
\cline{1-6}
$N_{exp} = 3$ fits to& 3/24 & 1 &0.135(2) &-0.0105(8) &  \\
 (1,loc) and (2,loc)&        & 2 &0.028(6) &$\;\,$0.100(8)&  \\
                  &        & 3 &0.030(9) &$\;\;$0.045(12)&  \\
\cline{1-6}
$N_{exp} = 3$ fits to& 7/24 & 1  &0.135(3) &-0.0104(8) &-0.0023(4)\\
 (1,loc) , (2,loc)&        & 2  &0.02(4) &$\;\,$0.14(8)&-0.01(3)  \\
and (3,loc)       &        & 3  &0.01(6) &-0.04(9)&$\;\,$0.11(2)  \\
\cline{1-6}
\end{tabular}
\end{center}
\caption{Examples of fit results for amplitudes $b(n_{sc},k)$}
\label{bamplitude}
\end{table}

\begin{table}
\begin{center}
\begin{tabular}{l|ccllc}
& $N_{exp}$ & $t_{min}/t_{max}$ &$\;\;aE_1\;\;$ & $\;\;aE_2\;\;$ & $Q$\\
\cline{1-6}
$2 \times 2$ matrix &2& 8/24 & 0.630(4) &0.82(3) & 0.15 \\
     fits      &    & 7/24 & 0.630(3)  & 0.81(2) & 0.17  \\
               &    & 6/24 & 0.632(5)  & 0.84(1) & 0.03 \\
               &    & 8/14 & 0.625(5)  & 0.83(1) & 0.19 \\
               &    & 7/14 & 0.626(4)  & 0.80(2) & 0.22 \\
\cline{1-6}
fits to (1,loc)& 2  & 8/24 & 0.630(6)  & 0.85(3) & 0.008\\
and (2,loc)    &    & 7/24 & 0.628(5)  & 0.82(2) & 0.008  \\
               &    & 8/14 & 0.628(10)  & 0.82(3) & 0.23  \\
               &    & 7/14 & 0.626(8)  & 0.81(2) & 0.35  \\
               & 3  & 7/14 & 0.629(16) &0.88(19) &0.22  \\
               &    & 4/14 & 0.626(9) &0.79(6)  &0.40  \\
\cline{1-6}
\end{tabular}
\end{center}
\caption{Examples of fits to ${^1P}_1$ correlations }
\label{pstates}
\end{table}

\vspace{.1in}
In Table~\ref{matrix} we give examples of $2 \times 2$ and $3 \times 3$
matrix fits (procedure 1) to ${^3S}_1$ levels for several $t_{min}/t_{max}$.
The results are consistent with Table~\ref{smeared-local} but with generally
 worse Q-values. For the $3 \times 3$ matrix fits one sees that the
signal for the second excited state has disappeared once $t_{min} > 6$.
The errors in Table~\ref{smeared-local} and ~\ref{matrix} are obtained by
the criterion that $\delta\chi^2 = 1$.  We have checked that bootstrap
errors agree with these errors to within $\pm$10\%.
 This indicates that the
statistical fluctuations in our correlations are close to being Gaussian.
In Fig.5 and Fig.6 we show ``effective amplitude'' plots (for $G_{meson}(t)\;
\cdot e^{E_1 \cdot t}$) corresponding to the $N_{exp}=3$,
$t_{min}/t_{max} = 5/24$ fit to two correlations in
 Table~\ref{smeared-local} and the
 $3 \times 3$ matrix fit with
$t_{min}/t_{max} = 7/22$ in Table~\ref{matrix}.
 One sees that most of the correlations are reproduced reasonably well
by our fits.

\renewcommand{\Diamond}{}

\setlength{\unitlength}{0.240900pt}

\begin{figure}
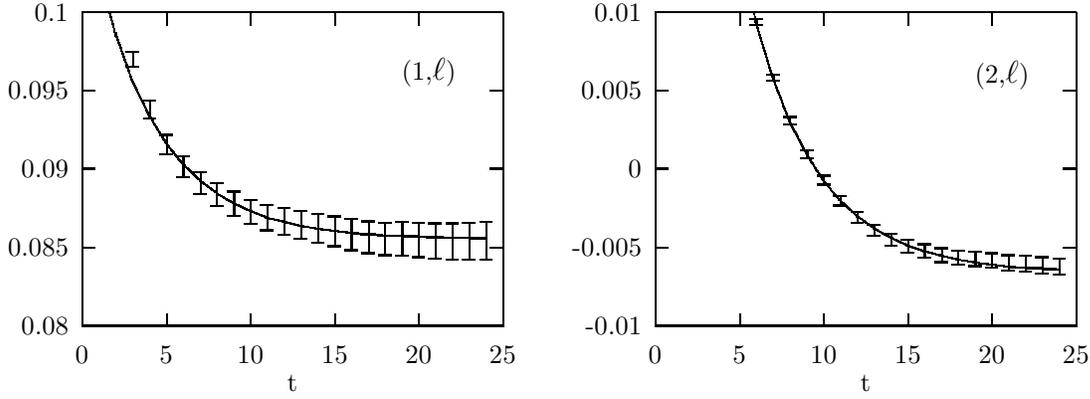


\noindent


\caption{$^3S_1$ Effective amplitudes
$G(t)\;\cdot e^{E_1 \cdot t} $
from three-exponential fits with \mbox{$t_{\rm min} = 5$},
\mbox{$t_{\rm max} = 24$}. }

\end{figure}

\renewcommand{\Diamond}{}

\setlength{\unitlength}{0.240900pt}

\begin{figure}
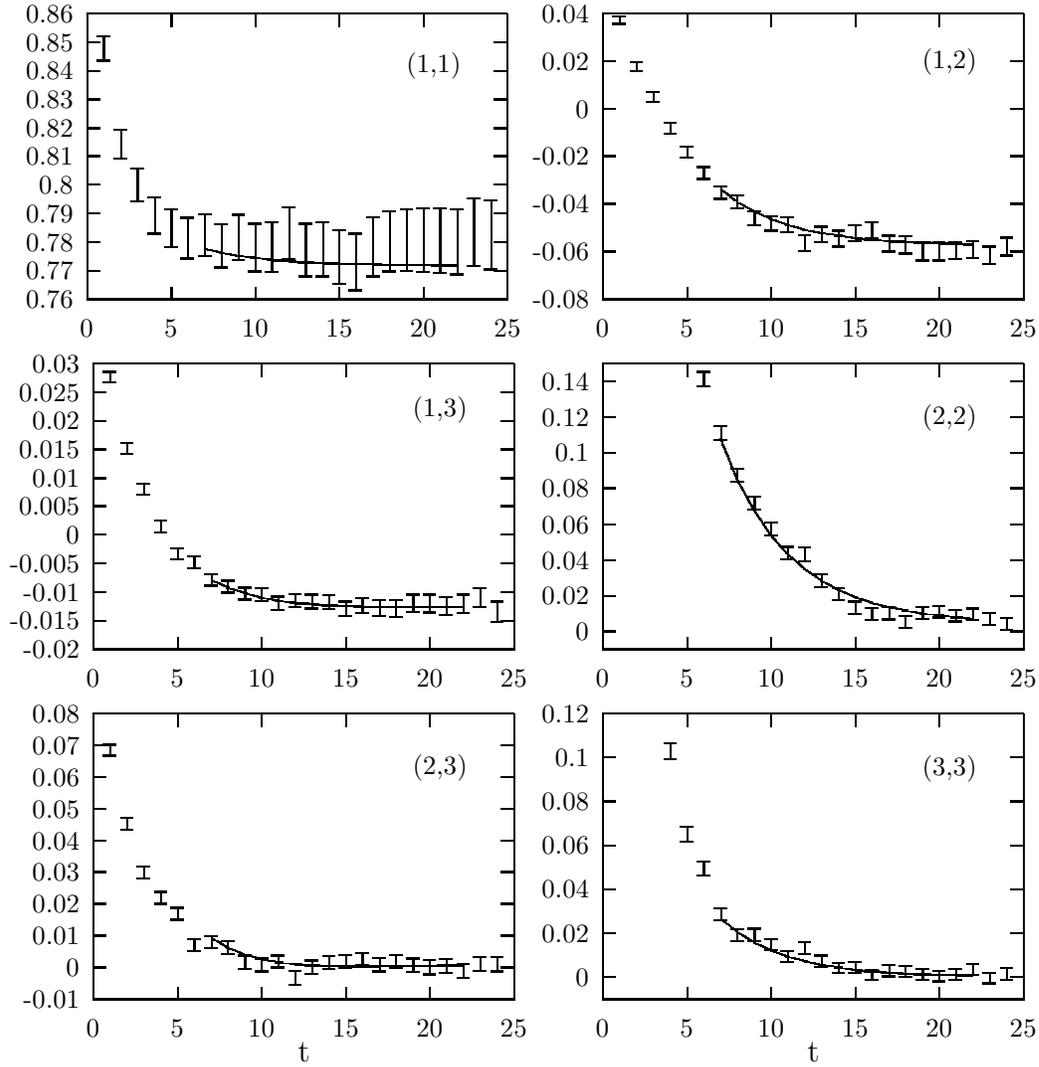


%
}
\put (530, -30) {t}
\put (1330, -30) {t}
}
\end{picture}

\caption{$^3S_1$ Effective amplitudes
$G(t)\;\cdot e^{E_1 \cdot t} $
for three by three matrix fits with
\mbox{$t_{\rm min} = 7$}, \mbox{$t_{\rm max} = 22$}.
Off-diagonal fits are averaged over source and sink.
}
\end{figure}

Information on how well our smearing functions are doing is contained
in the fitted amplitudes, the $a(n,k)$'s  and $b(n,k)$'s
 of  \eq{matansatz} and \eq{mcorfit}.  These measure the overlap between
our smearing functions and energy eigenstate wave functions.
 In Table~\ref{aamplitude}
 we show $a(n,k)$ for several matrix fits of Table~\ref{matrix} and
similarly in Table~\ref{bamplitude} we show $b(n,k)$ of smeared-local fits.
One sees for instance that with $n_{sc,sk} = 1$ , the amplitude for
the ground state (k = 1) dominates, whereas for $n_{sc,sk} = 2$ or
 $n_{sc,sk} = 3$ one has the largest overlap with, respectively
the first or second excited state (k=2 or k=3).
Our Richardson potential smearing functions are doing a reasonable job
in focusing correlations onto the right levels.  One might do better
by using our own simulations to provide good smearing functions.
 We are planning to do this in future simulations.

\vspace{.1in}
The multi-correlation multi-exponential fits described above were used to
obtain energies for the $\eta_b$, $\Upsilon$, the $h_b$ levels
and the singlet D-states.    The
$\eta_b$ fits are similar to those shown for $\Upsilon$ states.
 The $h_b$ and the D-state correlations are more noisy.  We give
examples of fits for the $h_b$ levels in Table~\ref{pstates}
 using  $2 \times 2$
matrix fits (with $(\nscsk)$ = (1,1),(1,2),(2,1) and (2,2))
and using simultaneous fits to $(\nscsk)$ = (1,loc) and
(2,loc).
We will give summaries of all our fits for different $\mbare$
below, where we also
convert dimensionless numbers into real energies in GeVs.  Here we go on to
describe fitting procedures used to obtain fine-structure splittings
between the $\chi_b$ states, the $\Upsilon-\eta_b$ splitting and the splitting
between states with zero and nonzero momenta.

\subsection{Fitting Procedures for Spin Splittings}
\vspace{.2in}
One test of the NRQCD effective action is to
see how well the fine
and hyperfine structure in the $b \bar{b}$ system can be reproduced. These
splittings will be
the ones most sensitive to the coefficients $c_i$ in the action and the
question is whether, with tadpole-improvement, tree-level
values of $c_i = 1$ are adequate.
The splittings between the $\chi_b$ states
are a few tens of MeV.  {}From Table~\ref{pstates}
 one sees that direct determination of
P-state levels have errors that are at best 0.005 - 0.010
 in dimensionless units.
With an inverse lattice spacing of $a^{-1} \simeq 2.4$ GeV (see next
section) this corresponds to an error of 12 to 24MeV
 and it seems
marginal whether we would be able to resolve fine structure if each P-level
were fit independently.
To get around this problem one can take advantage of the fact that
correlation functions for different mesons on the same configuration
can be highly correlated.  This allows for a direct fit to the mass
splitting between them.   We have employed two
methods for extracting splittings, $\delta  E = E^{A}_1 - E^{B}_1$,
from two very correlated
propagators for mesons, $A$ and $B$, having different quantum numbers.
 We call them the ``ratio'' method and the ``correlated-
$\delta  E$'' method.

\vspace{.1in}
\noindent
In the ratio method we take our 840 measurements for each meson $A$ and $B$
and create a jackknife ensemble of ratios of correlations. This is then
fit to a single exponential to determine $\delta E$.  In the correlated-
$\delta E$ method we pick a set of correlations for each meson and fit
the two sets simultaneously using

\begin{eqnarray}
G_{meson\;A}(n_{sc},loc;t) &=& \sum_{k=1}^{N_{exp}}\; c_A(n_{sc},k)\,
e^{-E_k^A \cdot t} \nl
G_{meson\;B}(n_{sc},loc;t) &=& c_B(n_{sc},1)\,e^{-(E_1^A+ \delta E) \cdot t}
\; + \;\sum_{k=2}^{N_{exp}}\; c_B(n_{sc},k)\,e^{-E_k^B \cdot t}
\end{eqnarray}
Of the two methods the second is more general, since it is straightforward
to include higher exponentials and handle spin splittings of excited
states.
We have estimated the effects of higher states on the ratio method by
fitting  the jackknifed ratio ensemble to

\be
Ratio(t) = A_1 e^{-\delta E \cdot t}\;{ { 1 + A_2e^{-\delta E_2 \cdot t}}
 \over {1 + A_3e^{-\delta E_2 \cdot t}}}
\ee
We did not find a signal for the correction terms
and the value for $\delta E$ was consistent
with those from naive one-exponential fits only with slightly larger errors.
  We give examples of fits
to the $\Upsilon$ - $\eta_b$ splitting and one of the
$\chi_b$ splittings in Tables~\ref{shyper} and ~\ref{phyper}.
For the ratio method we use (1,loc) correlations and for the
correlated-$\delta E$ method (1,loc) and (2,loc) correlations for each
meson.
 We see that it is possible
to get $\Upsilon - \eta_b$ splittings with 0.5MeV and $\chi_b$
splittings with 5 - 10MeV statistical errors (again using $a^{-1}$ of
2.4GeV).

\begin{table}
\begin{center}
\begin{tabular}{l|ccccc}
& $N_{exp}$ & $t_{min}/t_{max}$ & $a \delta E$ &  $Q$\\
\cline{1-5}
Ratio Method  & 1 & 18/24  &  0.0123(2)   & 0.21  \\
              & 1 & 16/24  &  0.0123(2)   & 0.21  \\
              & 1 & 12/24  &  0.0126(2)   & 0.09  \\
              & 1 & 8/24  &  0.0129(1)   & 0.005  \\
\cline{1-5}
correlated    & 2 &  8/24  &  0.0123(2)   & 0.39  \\
$\delta E$ fits&2 &  7/24  &  0.0123(2)   & 0.49  \\
              & 2 &  6/24  &  0.0122(2)   & 0.48  \\
\cline{1-5}
\end{tabular}
\end{center}
\caption{${^1S}_0 \; - \; {^3S}_1$ splitting }
\label{shyper}
\end{table}

\begin{table}
\begin{center}
\begin{tabular}{l|ccccc}
& $N_{exp}$ & $t_{min}/t_{max}$ & $a \delta E$ &  $Q$\\
\cline{1-5}
Ratio Method  & 1 & 10/24  &  0.021(4)   & 0.77  \\
              & 1 & 9/24  &  0.020(4)   & 0.79  \\
              & 1 & 8/24  &  0.017(3)   & 0.70  \\
              & 1 & 7/24  &  0.016(2)   & 0.76  \\
              & 1 & 6/24  &  0.017(2)   & 0.77  \\
\cline{1-5}
correlated    & 2 &  8/24  &  0.023(7)   & 0.52  \\
$\delta E$ fits&2 &  7/24  &  0.021(6)   & 0.55  \\
              & 2 &  6/24  &  0.018(5)   & 0.64  \\
\cline{1-5}
\end{tabular}
\end{center}
\caption{Example of P-state Fine Structure: ${^3P}_2 \; - \;
{^3P}_0$ splitting }
\label{phyper}
\end{table}

\vspace{.2in}
\subsection{Fit Results }
Using the fitting procedures of the previous subsections we obtained
energies for the $\Upsilon,\; \Upsilon', \; \Upsilon''$ levels,
the $h_b, \; h_b'$ levels, the $\Upsilon \, - \, \eta_b$ splitting
and for the splittings between the $\chi_b$ levels. For $a\mbare = 1.71$
we also have results for D-states (to date
we have only looked at averages over the five polarizations
of the ${^1D}_2$ state and leave more detailed studies of D-state fine
structure and mixing with S-states to future work).
  Our estimates
for the energies and splittings in dimensionless units are shown in
Table~\ref{summary}
 for several $\mbare$ values.
  For comparison we
also show results for data without relativistic and finite lattice spacing
corrections ($\delta H = 0$) and
for another set of data without  tadpole-improvement ($u_0 = 1$).
The errors in Table~\ref{summary} differ slightly from the purely statistical
errors of previous tables. In those tables one saw that
 the central values and statistical errors
depend sometimes on the fitting procedure used (and on $t_{min}/t_{max}$)
and we have tried to take that
into account in Table~\ref{summary}.  Only fits with good Q-values
( $Q \geq 0.2$ ) and
effective amplitude plots were included in these considerations.
  The reader may want to compare
 earlier tables with the $a \mbare = 1.71$ column in
Table~\ref{summary}.

\vspace{.1in}
There are several features worth noting in Table~\ref{summary}.  For instance,
 the splittings between $1{^3S}_1$ and $1{^1P}_1$ (spin averaged P-state)
 as well as between $2{^3S}_1$ and $1{^3S}_1$ levels are insensitive to
$\mbare$ within errors.  This is known to hold in the real world when one
compares splittings in the $J/\Psi$ and $\Upsilon$ systems. Also, for spin
averaged quantities the $\delta H = 0$ results are almost indistinguishable
from results with higher order corrections, except maybe for a slight
lowering of the 1S-level.  Considering next the spin dependent splittings,
one sees that the ${^3S}_1$-${^1S}_0$ splitting does depend on
 $\mbare$, as expected. More striking is the sensitivity of spin splittings
to tadpole-improvement.  The splittings are reduced by a factor of
$\sim {1 \over 2}$ without tadpole-improvement (i.e. if $u_0 = 1$).
As we shall see, the tadpole-improved results agree well with
experiment.
Tadpole-improvement of the lattice action appears crucial if
  one wants to work with tree-level values $c_i = 1$.  Otherwise spin
splittings are badly underestimated.

\begin{table}
\begin{center}
\begin{tabular}{c|lllll}
& $a\mbare$ = 1.71 & $\quad $1.8   &$\;\;$ 2.0  &
1.8 ($\delta H = 0$) &1.8 ($u_0 = 1$) \\
\cline{1-6}
$1{^3S}_1$ &0.4534(8)&0.4505(10)&0.444(1)&0.448(1)&1.097(1)  \\
$2{^3S}_1$ &0.695(10)&0.69(1)   &0.68(1)&0.69(1)&1.335(8)  \\
$3{^3S}_1$ &0.82(5)  &0.83(4)   &0.83(5)&0.84(5)& 1.48(3)  \\
$1{^1P}_1$ &0.626(8)&0.627(12) &0.619(10)&0.634(13)&1.27(1)  \\
$2{^1P}_1$ &0.81(3)  &0.80(4)   &0.79(4)&0.81(4)&1.43(3)  \\
$1{^1D}_2$ &0.76(3)  &&&&  \\
\cline{1-6}
Splittings: &&&&& \\
${^3S}_1\;-\;{^1S}_0$ &0.0123(2)&0.0116(2)&0.0106(2)&&0.0049(1)  \\
${^3P}_2\;-\;{^3P}_0$ &0.020(4)&0.021(4)&0.018(4)&&0.011(3)  \\
${^3P}_2\;-\;{^3P}_1$ &0.008(2)&0.008(3)&0.008(3)&&0.0047(14)  \\
\cline{1-6}
\end{tabular}
\end{center}
\caption{Fit results for dimensionless energies
and splittings, $aE$ and $a \delta E$.}
\label{summary}
\end{table}

\vspace{.2in}

\section{Comparisons with Experiment }

\vspace{.1in}
To compare our results with experiment we must
 convert from dimensionless lattice units to physical units
 by setting the scale $a^{-1}$.  Before doing so, it
is worthwhile reminding ourselves of systematic errors still contained
in our simulations.
  The largest source of systematic
errors, we believe, comes from the quenched approximation.
One expects,
among other things, corrections to both the 1S-1P and the 1S-2S splittings,
and inverse lattice spacings obtained by fitting these splittings
to experimental data should differ from each other. For instance,
a bootstrap estimate for the ratio of splittings, ($E(2{^3S}_1) -
E(1{^3S}_1)$)/($E(1{^1P}_1) - E(1{^3S}_1)$) $\equiv R$, gives R = 1.41(7)
for $a\mbare = 1.71$; R = 1.38(8) for $a\mbare = 1.8$; and
R = 1.35(8) for $a\mbare = 2.0$.  The experimental value for this ratio is
$R_{exp} = 1.28$ and one sees a possible 1 to 2 $\sigma$ deviation in
our simulations. Inverse lattice spacings obtained from matching the
S-P splitting to the Particle Data book will differ from that obtained
using the 1S-2S splittings by 1 or 2 $\sigma$.  Taking splittings from
Table~\ref{summary} that is indeed what we find.
  Work is underway to repeat
simulations of the $\Upsilon$ system with unquenched gauge configurations.
It will be interesting to see how the ratio of splittings will shift. {}From
potential models one can argue that the 1S-1P splitting will suffer a
larger change than the 1S-2S splitting, and that the ratio, R, defined
above will decrease as one goes from the quenched to unquenched world.
The other two major sources of systematic error in our calculation are
fitting errors and the finite lattice spacing and relativistic errors
in the NRQCD propagators.  Our experience with fitting indicates that
the results tend to move around by about a standard deviation when
different methods are used, in other words the systematic fitting
errors are probably about the same size as the statistical errors,
while we expect systematic errors in the propagators to contribute at
the 5 MeV level to the spectrum.

\vspace{.1in}
 In order to present dimensionful results we need to define one
global ( or ``average'') $a^{-1}$.  We do so by calculating a bootstrap
average based on the 1S-2S and 1S-1P splittings.  The results are
shown in Table~\ref{ainv}.  There we give $a^{-1}$'s, separately for
the two splittings plus the bootstrap values for averages and differences.
  We show
results for two different ensembles for the P-states, the first obtained
using smeared-local fits and the second based on matrix fits
( see Table~\ref{pstates} for spread in results due to fitting
procedure dependence)).
We use the bootstrap average plus its associated error to define
a global $a^{-1}$:

$$ a^{-1} = 2.4(1)GeV  $$

\vspace{.1in}
\noindent
Using the central value $a^{-1} = 2.4$GeV, one can
convert Table~\ref{summary} into physical numbers.
In Table~\ref{results} we do that for $a\mbare = 1.71$.
and compare with experiment. The agreement between simulation
results and experiment is excellent, with most entries agreeing within
$1 \sigma$.  It is hoped that once unquenched calculations are completed,
lattice QCD results will follow experiment
 even more closely.  Our experience with
the present quenched calculations tells
us that unquenching effects are small (this observation also follows from
potential model calculations)
and that high statistics data are
required to see them.
 The entries in Table~\ref{results} that do not have a corresponding
experimental number attached are predictions of the theory.
We predict $\Upsilon$ D-states
with center of mass at (10.20 $\pm$ 0.07 $\pm$ 0.03)GeV , where
the second error comes from uncertainties in $a^{-1}$.
  Our current
simulations give an $\eta_b$ state at (9.431 $\pm$ 0.005
 $\pm$ 0.001)GeV, where now the first (and dominant) error is due to
higher order relativistic and finite $a$ corrections not included in our
action and the second error is due to $a^{-1}$.
The $\eta_b$ energy is likely to change (decrease) when one goes beyond
the quenched approximation.  Given the small statistical errors on the
$\eta_b$ - $\Upsilon(1S)$ splitting
 this is one place where effects of quenching could be observable.

\vspace{.1in}
The spectrum results are shown in Figures~\ref{spect} and ~\ref{fs}
of section 1.  The P-state fine structure is measured relative to the
center of mass of the triplet states.  We first create bootstrap ensembles
of individual P-state energies.  We then calculate bootstrap estimates for
the center of mass and for energies relative to the center of mass.
In Fig.~\ref{fs} the energy of the ${^1P}_1$-state (the $h_b$) lies
slightly below the triplet center of mass.  We note, however, that
within our $\sim$5 MeV systematic errors the two levels are consistent
with each other.

\begin{table}
\begin{center}
\begin{tabular}{c|ll|llll}
           &from Table~\ref{summary} &&$\quad$Bootstrap& Results&& \\
$a\mbare$  & $a^{-1}(1S-2S)$&$a^{-1}(1S-1P)$
  & $a^{-1}(1S-2S)$&$a^{-1}(1S-1P)$& Average $a^{-1}$& $\Delta a^{-1}$ \\
\cline{1-7}
1.71  &2.33(11)  & 2.55(12)  & 2.34(8)  &  2.49(4)  &2.42(6)  & 0.15(7)  \\
      &          &          &          &  2.58(14) &2.46(9)  & 0.23(13) \\
      &&&&  \\
1.8   &2.35(11)  & 2.49(17) & 2.32(10) & 2.49(11)  & 2.41(9) & 0.17(10)  \\
      &          &          &          & 2.50(15)  & 2.41(10)& 0.18(15) \\
      &&&&  \\
2.0   &2.39(11)  & 2.51(15) & 2.39(12) &2.50(10) &  2.45(10) & 0.12(10)   \\
      &          &          &          &2.52(14)  &  2.45(11) &   0.13(14)\\
\cline{1-7}
\end{tabular}
\end{center}
\caption{$a^{-1}$ from 1S-2S and 1S-1P splittings }
\label{ainv}
\end{table}

\begin{table}
\begin{center}
\begin{tabular}{c|ll}
 & Simulation Results [GeV]  &  Experiment [GeV]  \\
\cline{1-3}
$2{^3S}_1\;-\;1{^3S}_1$ &0.580(26)  &  0.563  \\
$3{^3S}_1\;-\;1{^3S}_1$ &0.88(12)   &  0.895  \\
$1{^1P}_1\;-\;1{^3S}_1$ &0.414(22)  &  0.440  \\
$2{^1P}_1\;-\;1{^3S}_1$ &0.86(7)    &  0.800  \\
$1{^1D}_2\;-\;1{^3S}_1$ &0.74(7)    &  \\
\cline{1-3}
${^3S}_1\;-\;{^1S}_0$ &0.0295(5)  &  \\
${^3P}_2\;-\;{^3P}_0$ &0.048(10)  & 0.053  \\
${^3P}_2\;-\;{^3P}_1$ &0.019(5)   & 0.021  \\
${^3P}_{CM}-{^1P}_1$  &0.005(1)   &   \\
\cline{1-3}
\end{tabular}
\end{center}
\caption{NRQCD spectrum results
and comparison with experiment. We use $a^{-1}$ = 2.4GeV and
$a\mbare$ = 1.71.  Systematic errors due to quenching, higher order terms
in the NRQCD action and uncertainty in the scale $a^{-1}$ are not included.}
\label{results}
\end{table}

\vspace{.1in}

One quantity of phenomenological interest is the mesonic wave function at the
origin (i.e. zero
separation between quark and anti-quark), which for the $\Upsilon(mS)$ is
\begin{eqnarray}
 \Psi_m(0) \;  &=& \;
    <m| \, \sum_{\xv} \psid(\xv) \, \sigma_z \, \chi^\dagger(\xv) \, |0>, \nl
           \;  &=& \; <m|loc>_{^3S_1z}
\label{wavefuncdef}
\end{eqnarray}
where $|0>$ is the fock space vacuum, $<m|$ is the quantum state of an
$\Upsilon(mS)$ polarized in the $z$ direction, and $|loc>_{^3S_1z}$ is
the state created by our $loc$ smearing function in the ${^3S_1z}$ channel.
Examining \eq{matansatz}, we see that (in the
limit of an infinite number of exponentials in our fit ansatz)
$<m|loc>_{^3S_1z}$ measured in units of
$a^{-3/2}$ is just the fit parameter
$a(loc,m)$, which in turn is equal to $\sqrt{b(loc,m)}$.  We extract
$b(loc,m)$ by fitting simultaneously to $G_{{^3S}_1}(loc,loc;t)$
$G_{{^3S}_1}(1,loc;t)$, $G_{{^3S}_1}(2,loc;t)$ and
$G_{{^3S}_1}(3,loc;t)$.  These are fits of the same form as \eq{mcorfit},
but including $n_{sc} = loc$.
In addition to obtaining $a^{3/2}<m|loc>$ directly from our fits, it can be
obtained indirectly from our separate fits to $a$ and $b$,

\be
 a^{3/2}\,\Psi_m(0) = \frac{b(n_{sc},m)}{a(n_{sc},m)}
= a^{3/2}\,\frac{<n_{sc}|m>\,<m|loc>}{<n_{sc}|m>},
\label{indirect}
\ee
for any initial smearing function $n_{sc}$.  In Table~\ref{meswave}
we have tabulated  measurements of $a^{3/2}\,\Psi_m(0)$, from direct fits
to $b(loc,m)$ and from ratios of $b(n_{sc},m)$ and $a(n_{sc},m)$ taken
from Table~\ref{aamplitude} and Table~\ref{bamplitude}.  We also present
results for $\eta_b$.
 For ground states, values obtained using different initial smearing
functions are consistent. In general, we believe the diagonal entries in
Table~\ref{meswave} with $n_{sc} = m$ are the most reliable.  Good smearing
functions imply that off-diagonal amplitudes are severely suppressed.  Signal
to noise for the off-diagonal amplitudes $a(n,m)$ and $b(n,m)$ and for their
ratios is not as good as for the diagonal amplitudes.  The ``direct fit''
estimates for $\Psi(0)$ suffer from the fact that $b(loc,m)$ comes from
fitting the local-local meson correlation function.  This correlation has
many more exponentials contributing out to large t-values than is the case
for  smeared-local correlations (this is the reason why the local-local
correlation was not used in any of our fits to extract energies).  Hence,
it has proven difficult to extract the amplitude for any given excited
state accurately using the local-local correlation.  The indirect method
of \eq{indirect} gets around this problem and uses only smeared-local
and smeared-smeared correlations.

Care should be used in any attempt to relate these values of the wave function
to predictions of physical processes, such as the leptonic width of the
$\Upsilon$.
A complete calculation of such processes will involve large
($10$ - $20\%$ or more) corrections in relating the lattice current appearing
in \eq{wavefuncdef}
to continuum currents. These corrections arise both from relativistic
(coming from the small components of the heavy Dirac spinors)
and renormalization
effects.  In addition, $|\Psi(0)|^2$ scales like $a^{-3}$; meaning that
lattice spacing errors in the dimensionful value  of $|\Psi(0)|^2$
will be of order $13\%$.
Quenching effects are likely to induce an
additional systematic error
 on the simulation results, which we have not attempted to estimate
quantitatively.
 If one nonetheless goes ahead and combines the numbers in
 Table~\ref{meswave} with $a^{-1} = 2.4(1)$GeV and with the standard
leading order Van Royen-Weisskopf formula for a vector meson leptonic
width \cite{perkins},
one finds  for the $\Upsilon(1S)$,
$\Gamma_{ee}$ = (1.09 $\pm$ 0.03 $\pm$ 0.14)keV.
The experimental value is
$\Gamma_{ee}^{(exp.)}(\Upsilon)$ = (1.34 $\pm$ 0.04)keV.  In the simulation
result
we only quote two errors corresponding respectively to statistical errors
and  $a^{-1}$ uncertainties. Quenching errors and relativistic
and matching corrections (corrections that are expected to dominate)
 have not been included.
Unquenching should increase the theoretical estimate for $\Gamma_{ee}$ by
enhancing $|\Psi(0)|^2$.
Using the diagonal entries in Table~\ref{meswave},  one finds for the
$\Upsilon$ 2S and 3S, $\Gamma_{ee}(2S)$ = (0.66 $\pm$ 0.10 $\pm$ 0.09)keV
and
 $\Gamma_{ee}(3S)$ = (0.75 $\pm$ 0.37 $\pm$ 0.10)keV.  The experimental
numbers are respectively (using the $\mu^+ \mu^-$ branching ratios),
$\Gamma_{ee}^{(exp.)}(2S)$ = (0.56 $\pm$ 0.09)keV and
$\Gamma_{ee}^{(exp.)}(3S)$ = (0.44 $\pm$ 0.04)keV.
When one compares  $\Psi(0)$ given in Table~\ref{meswave} for the ground and
excited states
 one finds that the ratios $\Psi_{2S}(0)/\Psi_{1S}(0)$ and
$\Psi_{3S}(0)/\Psi_{1S}(0)$ are larger  than
one would expect from lattice potential calculations.
Our excited
state leptonic widths lie above the experimental numbers
( although consistent within large errors ), whereas for
$\Upsilon(1S)$ the simulation result lies below experiment.  Once
 corrections to the present calculation are included
( e.g. relativistic, quantum loop,
lattice-continuum matching, unquenching),  it will be important to
monitor and understand how and whether these
 discrepancies go away.  At the same time, one also needs to
reduce the statistical errors, particularly for the excited
states.  We hope
to be reporting on more accurate results in the future.

\begin{table}
\begin{center}
\begin{tabular}{c|l|lll}
$\qquad$Meson$\qquad$ &Direct Fit&$\qquad
\qquad$ Using& $b(n_{sc},m)/a(n_{sc},m)$&\\
$\quad (mS)$        &to $b(loc,m)$ &$n_{sc}=1$ & $n_{sc}=2$ & $n_{sc}=3$ \\
\cline{1-5}
$\Upsilon(1S)$&$\;\, .153(2)$&$\;\, .154(2)$ &$\;\, .162(12)$
 &$\;\, .17(2)$\\
$\Upsilon(2S)$&$\;\,  .147(10)$&$\;\, .16(3) $
 &$\;\, .127(10)$ &$\;\,----$\\
$\Upsilon(3S)$&$\;\,----$&$\;\, .32(10)$ &$\;\,----  $ &$\;\, .14(4)$\\
\cline{1-5}
$\eta_b(1S)$  &$\;\,.161(2) $ &$\;\,.162(2)$ &$\;\,.184(26)$ &$\;\,---- $\\
$\eta_b(2S)$  &$\;\,.147(9)$ &$\;\,.17(4)$ &$\;\,.132(10)$ &$\;\,----$\\
$\eta_b(3S)$  &$\;\,---- $ &$\;\,----$ &$\;\,----$ &$\;\,.13(2)$\\
\cline{1-5}
\end{tabular}
\end{center}
\caption{Mesonic wave function at the origin, $a^{3/2}\,\Psi(0)$.
 A dashed line means
 that no signal could be extracted.}
\label{meswave}
\end{table}

\vspace{.2in}

\section{$\Upsilon$ Mass and Lorentz Invariance}

As we have shown, NRQCD simulations gives accurate results for the splittings
between $b\overline b$~states. These simulations can also be used to
compute the full mass of the~$\Upsilon$. We use the~$\Upsilon$ mass to
determine the correct quark mass for the simulation.
We have investigated two different
methods for computing the $\Upsilon$~mass.

The first method is to add twice the renormalized mass of the quark to the
nonrelativistic energy~$E_{\rm NR}(\Upsilon)$ obtained from the
simulation:
\be
M_{\Upsilon} = 2 ( Z_m \mbare - E_0 ) +  E_{NR}(\Upsilon).
\label{pert}
\ee
Here $Z_m$ and $E_0$ are renormalizations that are computed using
perturbation theory \cite{M_b,colin}. Our results are shown in
Table~\ref{mrest}. The uncertainties in this procedure are due to
uncalculated $\order(\alpha_s^2)$~corrections in the perturbative expansions
of the renormalizations, and to uncertainties in~$a^{-1}$.
\begin{table}
\begin{center}
\begin{tabular}{c|ccc|c|c}
$a\mbare$ & $a\, E_{NR}$ & $Z_m$ & $a\,E_0$ & $a\, M_\Upsilon$ & $M_\Upsilon$
(GeV) \\ \hline
1.71 & 0.453(1) & 1.20(4) & 0.32(6) & 3.92(18) & 9.4(6) \\
1.8 & 0.451(1) & 1.18(4) & 0.31(6) & 4.08(18) & 9.8(6) \\
2.0 & 0.444(1) & 1.16(3) & 0.30(6) & 4.48(17) & 10.8(6)
\end{tabular}
\end{center}
\caption{The $\Upsilon$~mass as determined from the
$\Upsilon$'s~nonrelativistic rest energy in NRQCD. The last column assumes
$\ainv = 2.4(1)$.}
\label{mrest}
\end{table}

The second method is to compute the nonrelativistic energy
for~$\Upsilon$'s with nonzero three momenta. We did this for momenta
$\pv =$~(0,0,1), (0,1,1), (1,1,1) and (0,0,2) in units of~$4\pi/16a$,
and fit the resulting energies to two different parameterizations:
\be
E_\Upsilon(\pv) - E_{NR}(\Upsilon) = {\pv^2 \over {2\,M_{\kin}
}}  - C_1 {{(\pv^2)^2} \over {8 M_\kin^3}}
\ee
and
\be
E_{\Upsilon}(\pv) -  E_{NR}(\Upsilon)  =  {\pv^2 \over {2\,M_{\kin}
}}
   - { 1 \over {8 M_\kin^3}} \left\{(\pv^2)^2
\; + \; C_2 \sum_{k=1,2,3} p_k^4\right\}.
\ee
In a Lorentz invariant theory, the kinetic mass~$M_\kin$ is equal to
the rest mass of the upsilon; in our simulations this should be true
up to corrections of order~$v^4$, except in the $\delta H=0$ theory
where the errors should be order~$v^2$. The $(\pv^2)^2$ terms in our fits
test for higher-order relativistic effects; we expect $C_1 = 1$ up to
corrections of order~$\order(v^2,a^2)$, except in the $\delta H=0$
theory where this parameter should almost vanish.
The $p_k^4$ term in the second fit
tests for contributions that are not rotationally invariant; we
expect $C_2=0$
up to corrections of $\order(v^2,a^2)$. These expectations are
confirmed by our results which are shown in Table~\ref{mkin}.
\begin{table}
\begin{center}
\begin{tabular}{c|ccccc}
$a\mbare$  & $aM_{\kin}$   & $C_1$  &  $C_2$  & Q  &
$M_\kin$ (GeV)\\
\cline{1-6}
1.71  & 3.94(3)  & 1.0(3)  &        &  0.16  &  9.5(4)  \\
      & 3.94(3)  &           &0.20(37)  &  0.19 &  \\
      &&&&&  \\
1.8   & 4.09(3)  & 1.1(3)  &        &  0.60  &  9.8(4)   \\
      & 4.09(3)  &           &0.20(35)&  0.62  &             \\
      &&&&&  \\
2.0   & 4.48(4)  & 1.2(3)  &        &  0.64  &  10.8(4)  \\
      & 4.48(4)  &           &0.24(36)&  0.67  &             \\
      &&&&&  \\
1.8   & 3.82(3)  & 0.23(32) &        &  0.06 &    9.2(4)        \\
($\delta H = 0$)&3.83(3) &     &-0.45(41)& 0.006  &   \\
      &&&&&  \\
1.8   & 4.96(3)  & 1.4(3) &        &  0.11 &            \\
($u_0 = 1$)&4.97(3) &     &0.60(35)& 0.18  &   \\
\cline{1-6}
\end{tabular}
\end{center}
\caption{The $\Upsilon$ mass determined from its kinetic
mass~$M_{\kin}$ using two different fits. The parameters
$C_1$ and $C_2$ are explained in the text. The last column assumes
$\ainv = 2.4(1)$.}
\label{mkin}
\end{table}

Our two methods give identical $\Upsilon$~masses in lattice units to
better than~1\%. Assuming $\ainv=2.4(1)$, our
results indicate that the correct bare mass for a $b$~quark is
$\mbare = 1.7(1)/a$; among our simulations, the set
with~$a\,\mbare=1.71$ is the best.

The close agreement between our two determinations is striking confirmation
of the partial restoration of Lorentz invariance due to the correction
terms~$\delta H$ that we include in the quark
evolution equation. The kinetic
mass is quite sensitive to these correction terms; it is too small
by almost 10\% when we set~$\delta H = 0$. This is what we expect
since $M_\kin$ equals
twice the quark mass in a nonrelativistic theory,
while in a relativistic theory the binding energy contributes as
well. Note that tadpole improvement is also essential; without
it, $M_\kin$ is too large by almost 25\%.

Each of the two methods we use to determine the $\Upsilon$~mass has
its strengths and  weaknesses. Using~$M_\kin$ is attractive because
there is no need for perturbative calculations, but it is only
accurate if the $\delta H$~correction terms are included in the quark
propagators. The rest energy of the meson is quite insensitive
to~$\delta H$, but can only be used when combined with perturbative
results for the quark mass renormalizations. Ideally one uses both
methods and compares results, as we have done here.


\vspace{.4in}

\section{Summary}
\noindent
We have investigated the $\Upsilon$ system using nonrelativistic lattice
QCD (NRQCD) and find that one can successfully reproduce
 the general features of the known $\Upsilon$ spectrum,
 including P-state fine structure and several excited states.
We also make predictions for D-states and for $\eta_b$.
By reducing the statistical errors in the splittings by a modest amount,
one should, in the future,
 be able to study effects of quenching in a quantitative way.

  We worked with an action
correct through $\order(M_b \, v^4)$ and whose link variables had been
divided by $u_0$, the fourth root of the plaquette value, i.e.
with a tadpole-improved action.  Because of tadpole-improvement we were able
to get good results (e.g. the
 correct P-state fine structure splittings) with tree-level coefficients
for the correction terms in the NRQCD action.  Our program
should be viewed as a demonstration of a successful implementation of
perturbative improvement of a lattice action.  Our results underscore
the utility of tree-level improvement  and the crucial importance of
tadpole-improvement.

We have put considerable effort into developing fitting procedures
that allow us to get not just ground state energies
and energy splittings between levels with
different quantum numbers, but also (radially) excited energy levels.
We find that it is important to fit several correlations simultaneously,
and do multi-exponential multi-correlation fits.

Our calculations have systematic errors at the 5MeV level coming from
higher order relativistic and finite lattice spacing effects. The largest
source of uncertainty comes, however, from the  quenched approximation.
  Calculations involving
dynamical configurations are underway, and we will to be reporting on them
 soon.  Quenched $c\bar{c}$ spectrum results will also appear shortly.

\newpage
\vspace{.4in}

\centerline{ \underline{ \bf{Acknowledgements} }}

\vspace{.1in}
\noindent
This work was supported in part by grants from the U.S.Department of
Energy (DE-FC05-85ER250000, DE-FG05-92ER40742, DE-FG02-91ER40690), from
the National Science Foundation and from the UK SERC.  C.~Davies thanks
the physics department of the Ohio State University for visitor support
 and hospitality while part of this work was being completed.
  The numerical calculations were
carried out at the Ohio Supercomputer Center.  We thank Greg Kilcup for
making his configurations available to us.  Finally, we wish to
acknowledge fruitful discussions with Aida El-Khadra, Paul Mackenzie,
 Colin Morningstar and Beth Thacker.


\begin{thebibliography}{8}
\bibitem{previous}
The $\Upsilon$ spectrum based on our $a\mbare = 1.8$ data was presented in
 \cite {poster}.  We have now completed the analysis for several $a\mbare$
 values. The figures shown here are for $a\mbare = 1.71$, the
value of $a\mbare$ that gives the best absolute $\Upsilon(1S)$ mass.
\end{thebibliography}
\end{document}